# Identification of 45 New Neutron-Rich Isotopes Produced by In-Flight Fission of a $^{238}$U Beam at 345 MeV/nucleon

Tetsuya OHNISHI, Toshiyuki KUBO[*], Kensuke KUSAKA, Atsushi YOSHIDA, Koichi YOSHIDA, Masao OHTAKE, Naoki FUKUDA, Hiroyuki TAKEDA, Daisuke KAMEDA, Kanenobu TANAKA, Naohito INABE, Yoshiyuki YANAGISAWA, Yasuyuki GONO, Hiroshi WATANABE, Hideaki OTSU, Hidetada BABA, Takashi ICHIHARA, Yoshitaka YAMAGUCHI, Maya TAKECHI, Shunji NISHIMURA, Hideki UENO, Akihiro YOSHIMI, Hiroyoshi SAKURAI, Tohru MOTOBAYASHI, Taro NAKAO[1], Yutaka MIZOI[2], Masafumi MATSUSHITA[3], Kazuo IEKI[3], Nobuyuki KOBAYASHI[4], Kana TANAKA[4], Yosuke KAWADA[4], Naoki TANAKA[4], Shigeki DEGUCHI[4], Yoshiteru SATOU[4], Yosuke KONDO[4], Takashi NAKAMURA[4], Kenta YOSHINAGA[5], Chihiro ISHII[5], Hideakira YOSHII[5], Yuki MIYASHITA[5], Nobuya UEMATSU[5], Yasutsugu SHIRAKI[5], Toshiyuki SUMIKAMA[5], Junsei CHIBA[5], Eiji IDEGUCHI[6], Akito SAITO[6], Takayuki YAMAGUCHI[7], Isao HACHIUMA[7], Takeshi SUZUKI[7], Tetsuaki MORIGUCHI[8], Akira OZAWA[8], Takashi OHTSUBO[9], Michael A. FAMIANO[10], Hans GEISSEL[11], Anthony S. NETTLETON[12], Oleg B. TARASOV[12], Daniel P. BAZIN[12], Bradley M. SHERRILL[12], Shashikant L. MANIKONDA[13], and Jerry A. NOLEN[13]

*RIKEN Nishina Center, RIKEN, 2-1 Hirosawa, Wako, Saitama 351-0198*
[1]*Department of Physics, University of Tokyo, 7-3-1 Hongo, Bunkyo-ku, Tokyo 113-0033*
[2]*Department of Engineering Science, Osaka Electro-Communication University, 18-8 Hatsucho, Neyagawa, Osaka 572-8530*
[3]*Department of Physics, Rikkyo University, 3-34-1 Nishi-Ikebukuro, Toshima-ku, Tokyo 171-8501*
[4]*Department of Physics, Tokyo Institute of Technology, 2-12-1 Ookayama, Meguro-ku, Tokyo 152-8551*
[5]*Faculty of Science and Technology, Tokyo University of Science, 2461 Yamazaki, Noda, Chiba 278-8510*
[6]*Center for Nuclear Study, University of Tokyo, 2-1 Hirosawa, Wako, Saitama 351-0198*
[7]*Department of Physics, Saitama University, 255 Shimo-Okubo, Sakura-ku, Saitama City, Saitama 338-8570*
[8]*Institute of Physics, University of Tsukuba, 1-1-1 Ten'noudai, Tsukuba, Ibaraki 305-8571*

[9]*Institute of Physics, Niigata University, 8050 Ikarashi 2-no-cho, Nishi-ku, Niigata 950-2181*

[10]*Department of Physics, Western Michigan University (WMU), 1903 W. Michigan Avenue, Kalamazoo, Michigan 49008-5252, U.S.A.*

[11]*Gesellschaft fuer Schwerionenforshung (GSI) mbH, 1 Planckstr, Darmstadt 64291, Germany*

[12]*National Superconducting Cyclotron Laboratory(NSCL), Michigan State University (MSU), 1 Cyclotron, East Lansing, Michigan 48824-1321, U.S.A.*

[13]*Argonne National Laboratory (ANL), 9700 S. Cass Avenue, Argonne, Illinois 60439, U.S.A.*

A search for new isotopes using in-flight fission of a 345 MeV/nucleon $^{238}$U beam has been carried out at the RI Beam Factory at the RIKEN Nishina Center. Fission fragments were analyzed and identified by using the superconducting in-flight separator BigRIPS. We observed 45 new neutron-rich isotopes: $^{71}$Mn, $^{73,74}$Fe, $^{76}$Co, $^{79}$Ni, $^{81,82}$Cu, $^{84,85}$Zn, $^{87}$Ga, $^{90}$Ge, $^{95}$Se, $^{98}$Br, $^{101}$Kr, $^{103}$Rb, $^{106,107}$Sr, $^{108,109}$Y, $^{111,112}$Zr, $^{114,115}$Nb, $^{115,116,117}$Mo, $^{119,120}$Tc, $^{121,122,123,124}$Ru, $^{123,124,125,126}$Rh, $^{127,128}$Pd, $^{133}$Cd, $^{138}$Sn, $^{140}$Sb, $^{143}$Te, $^{145}$I, $^{148}$Xe, and $^{152}$Ba.

Since the pioneering production of radioactive isotope (RI) beams in the 1980s,[1] studies of exotic nuclei far from stability have been attracting much attention. Neutron-rich exotic nuclei are of particular interest, because new phenomena such as neutron halos, neutron skins, and modifications of shell structure have been discovered.[2-5] Furthermore these neutron-rich nuclei are important in relation to astrophysical interests,[6] because many of them play a role in the astrophysical r-process.[7] To make further advances in nuclear science and nuclear astrophysics, it is essential to expand the region of accessible exotic nuclei towards the neutron drip-line. In-flight fission of a uranium beam is known to be an excellent mechanism for this

---

E-mail: kubo@ribf.riken.jp

purpose, having large production cross sections for neutron-rich exotic nuclei.[8]

In 2007, at the RIKEN Nishina Center, a new-generation RI beam facility called the RI Beam Factory (RIBF)[9] became operational, in which the superconducting in-flight separator BigRIPS[10,11] has been used for the production of RI beams. The BigRIPS separator is designed as a two-stage separator with large acceptance, so that excellent features of in-flight fission can be exploited. In May 2007, right after the commissioning of the BigRIPS separator, we performed an experiment to search for new isotopes using in-flight fission of a 345 MeV/nucleon $^{238}$U beam, aiming to expand the frontier of accessible neutron-rich exotic nuclei. Even though the uranium beam intensity was low, we observed the new neutron-rich isotopes $^{125}$Pd and $^{126}$Pd, demonstrating the capability of the BigRIPS separator.[11] In November 2008, we revisited the experiment with upgraded beam intensity. The measurement was carried out using three different settings of the separator, each targeting new isotopes in the region with atomic numbers around 30, 40, and 50, respectively, and in total we observed 45 new neutron-rich isotopes. In this letter we report on the identification of these new isotopes that were produced for the first time using the BigRIPS separator.

The experiment was performed with a $^{238}$U$^{86+}$ beam at 345 MeV/nucleon. The beam intensity was ~0.22 particle nA (pnA) on average. The experimental method was essentially the same as we used in 2007.[11] The first stage of the BigRIPS separator was used to collect and separate fission fragments, while the second stage served as a spectrometer for particle identification (PID). An achromatic energy degrader was used in the first stage for selection of a range of isotopes to be measured. If further purification was needed, another degrader was used in the second stage. The PID was based on the $\Delta E$-TOF-$B\rho$ method, in which the energy loss ($\Delta E$), time of flight (TOF), and magnetic rigidity ($B\rho$) were measured to deduce the atomic number ($Z$) and the mass-to-charge ratio ($A/Q$) of fragments.

The experimental conditions were determined based on detailed simulations using the code LISE++ (version 8.4.1).[12] Table 1 summarizes the three separator settings used in the experiment. We refer to them as G1, G2, and G3, respectively. The angular acceptance of the separator was assumed to be the design values: horizontally ±40 mr and vertically ±50 mr,[10] while the momentum acceptance was set to ±3% by using slits at the F1 dispersive focus at the mid point of the first stage. The slits at the F2 achromatic focus at the exit of the first stage determined transmitted isotopes. (See Fig. 1 of ref. 11 for the detailed configuration of the BigRIPS separator.) The target material chosen for G1 and G2 was beryllium because the so-called abrasion-fission (AF) process is more favorable for the production of isotopes in the $Z$~30 to ~40 regions. On the other hand, we chose a lead target for G3, because the Coulomb-fission (CF) process that leads to asymmetric fission is more favorable for the production of isotopes in the Z~50 region. The code LISE++ includes both the AF and CF models of the fission process.[12] The $B\rho$ settings were chosen to select the high-momentum side of the distributions even for new isotopes, so that the overall count rate of fragments might not be too high for our data acquisition system. In the case of G3, we used an aluminum foil directly behind the target to increase the fraction of fully stripped ions, and inserted the second degrader at the F5 dispersive focus at the mid point of the second stage to improve the purity.

The TOF was measured between two thin plastic scintillation counters (PLs) placed at the F3 and F7 achromatic foci, which are located at the beginning and end of the second stage, respectively. The $\Delta E$ was measured at F7 using a multi-sampling ionization chamber (MUSIC).[13] Six energy-loss signals obtained from the MUSIC were averaged and used for the $\Delta E$ measurement. The $B\rho$ measurement was made by trajectory reconstruction not only in the first half but also in the second half of the second stage. For the trajectory reconstruction, the positions and angles of fragments were measured at the F3, F5, and F7 foci by using two sets of position-sensitive parallel plate avalanche counters (PPACs)[14] placed at the respective foci. First-order ion-optical transfer maps obtained experimentally and second-order transfer maps determined empirically were used for the trajectory reconstruction. The twofold $B\rho$ measurement was needed to deduce the $A/Q$ value of fragments in combination with the TOF measurement, because the fragments were slowed down at F5 due to the PPAC detectors and the energy degrader. The PPACs were also used for additional TOF measurements. The methods of calibration of the TOF and $\Delta E$ measurements were described in ref. 11. After the BigRIPS separator, the fragments were transported to the end of the ZeroDegree spectrometer,[15] the F11 focus,[11] where the PID was confirmed by detecting delayed $\gamma$-rays from isomeric states by using three clover-type Ge detectors. In the case of G1, we used another MUSIC placed at F11 for the $\Delta E$ measurement.

Inconsistent events were excluded by checking phase space profiles as well as beam spot profiles of fragments, consistency of fragment trajectories, and various correlation plots made of pulse-height signals and timing signals in the PL, PPAC and MUSIC detectors. We compared the two $B\rho$ measurements to reject inconsistent events. This also allowed us to determine whether or not the charge state of fragments changed at F5. The two TOF measurements were also compared to exclude inconsistent events.

Figures 1 (a)-(c) show the PID plots of $Z$ versus $A/Q$ for the three settings. The events that changed their charge states at F5 are not included in the figures. The relative root mean square (rms) $Z$ resolution and the relative rms $A/Q$ resolution achieved were typically 0.56% and 0.056% for G1, 0.57% and 0.035% for G2, and 0.42% and 0.041% for G3. These values are the estimates for Zn, Zr, and Sn isotopes, respectively. The lower panels of Fig. 1 show the PID plots enlarged around the regions of new isotopes. The red solid lines indicate the limit of previously identified isotopes. Figures 2 (a)-(c) show the projected one-dimensional $A/Q$ spectra: (a) for G1 (Ca to Y isotopes), (b) for G2 (Se to In isotopes), and (c) for G3 (Rh to Ba isotopes). The $A/Q$ spectra were obtained by gating the PID plots with $Z$ gates set between $Z \pm \sigma_Z$, where $\sigma_Z$ represents the absolute rms $Z$ resolution. The gates are shown as dotted lines in the lower panel of Fig. 1. Thanks to the excellent resolution in $A/Q$, the peaks for fully stripped ($Q=Z$), hydrogen-like ($Q=Z-1$) and helium-like ($Q=Z-2$) ions are well separated from each other, so that new isotopes can be clearly identified. In total we have produced and identified the following 45 new isotopes: $^{71}$Mn, $^{73,74}$Fe, $^{76}$Co, $^{79}$Ni, $^{81,82}$Cu, $^{84,85}$Zn, $^{87}$Ga, $^{90}$Ge, $^{95}$Se, $^{98}$Br, $^{101}$Kr, $^{103}$Rb, $^{106,107}$Sr, $^{108,109}$Y, $^{111,112}$Zr, $^{114,115}$Nb, $^{115,116,117}$Mo, $^{119,120}$Tc, $^{121,122,123,124}$Ru, $^{123,124,125,126}$Rh, $^{127,128}$Pd, $^{133}$Cd, $^{138}$Sn, $^{140}$Sb, $^{143}$Te, $^{145}$I, $^{148}$Xe, and $^{152}$Ba. They are labeled by their mass numbers in Fig. 2, and listed in Table 2 along with production yields.

The absolute rms $A/Q$ resolution ($\sigma_{A/Q}$) is much better than the peak separation in the $A/Q$ spectra, allowing the clear identification of the new isotopes. For instance, the most severe case is the peak separation between $^{84}$Zn$^{30+}$ (new isotope) and $^{81}$Zn$^{29+}$ (neighboring hydrogen-like ions) in the G1 setting, which is $6.0 \times 10^{-3}$, corresponding to $3.8 \sigma_{A/Q}$. The best case is the separation between $^{119}$Tc$^{43+}$ (new isotope) and $^{116}$Tc$^{42+}$ in the G2 setting, which is $5.4 \times 10^{-3}$, corresponding to $6.8 \sigma_{A/Q}$. The centroids of the observed $A/Q$ peaks agree well with the calculation using mass values, thanks to the accuracy of calibration. The deviation is within $1.0 \times 10^{-3}$ in terms of the absolute $A/Q$ value, which is almost the same as the $\sigma_{A/Q}$. This also helped the clear identification.

Because we detected only a few events for some of the new isotopes in the expected region of the $A/Q$ spectrum ($\pm 2\sigma_{A/Q}$), we made a significance test[16] to determine if these events could have originated from the tails of neighboring hydrogen-like peaks. The probability that all events come from the tails was evaluated statistically using the achieved resolution values of $A/Q$ and $Z$ and assuming a Poisson distribution. The result is listed in Table 2 as p-value,[16] which gives the probability of misidentification as a new isotope. In the test we concluded that the observation was from the identification of a new isotope, if the p-value is smaller than 1% (significance level). Thanks to the achieved *A/Q* and *Z* resolution, it was concluded that we observed the new isotopes $^{112}$Zr, $^{124}$Ru, $^{126}$Rh and $^{148}$Xe even though they had only one count. However, in a few cases such as $^{99}$Br (G2, p-value = 1.1%) and $^{129}$Pd (G2, 1.7%), we could not exclude the possibility of the neighboring-peak origin because the p-value is greater than 1%. Furthermore we estimated the rate of random background events that were uniformly distributed in the new isotope region of the PID plot. For all the three settings, the estimates are on the order of 0.01 counts per unit peak area, which are much smaller than the least counts of observed new isotopes. Here the unit peak area is defined as $|Z-Z_0| \leq \sigma_Z$ and $|A/Q-(A/Q)_0| \leq 2\sigma_{A/Q}$, where $Z_0$ and $(A/Q)_0$ represent the peak centriod in the *Z* versus *A/Q* plot.

Figure 3 shows the measured production rates along with the predictions from the LISE++ simulations. Note that the rates shown are for events corresponding to fully stripped ions throughout the separator. The LISE++ simulations were made using the AF model for the beryllium target (G1 and G2), while for the lead target (G3) the CF model was used in combination with the AF model. The AF model for the beryllium target relies on the so-called three excitation energy model in which three nuclei, $^{236}$U, $^{226}$Th and $^{220}$Ra, are chosen to represent all the fissile nuclei created in the abrasion-ablation stage, then followed by fission fragment distributions. For the lead target, the three representative nuclei were $^{238}$U, $^{231}$Th, and $^{215}$Po, and the CF model was included to simulate the Coulomb excitation of the $^{238}$U nucleus. The details are given in the LISE++ manual and the recommended fission parameters therein were used.[12] The simulations were made in the Monte Carlo mode in which secondary reactions in the target and degrader materials are not included. The measured rates are fairly well reproduced by the LISE++ predictions. The systematics of the measured rates as well as the reproduction by the LISE++ simulations supports the identification of new isotopes.

We estimated experimental production cross sections based on the ratio of the measured production rate to the predicted production rate and the predicted cross sections given by the code LISE++. The estimates are given in Table 2 for the new isotopes. We estimate that our method for determination of the cross sections has systematic errors of ~50%, ~40%, and ~30% for the G1, G2, and G3 settings, respectively. The errors originate from the evaluation of the transmission efficiency and the determination of the beam intensity. The dominant error is statistical for isotopes with low count rates.

In summary, we have conducted a search for new isotopes using in-flight fission of a $^{238}$U beam at 345 MeV/nucleon, and observed 45 new isotopes over a wide range of atomic numbers. Figure 4 shows the newly discovered isotopes on a nuclear chart that includes an estimated r-process path based on the KTUY mass formula.[17] For Pd, we observed the new isotopes $^{127}$Pd and $^{128}$Pd, and reached the r-process waiting point at the $N$=82 neutron magic number.[18] The present results demonstrate the significant potential of the RIBF, which promises to vastly expand the accessible region of exotic nuclei, moving towards the drip-line as the primary beam intensity increases over time. As the production rates increase, more detailed information on such important neutron-rich nuclei, for example, decay properties, shapes and single particle structure,[18] can be studied via a combination of the BigRIPS separator and the ZeroDegree spectrometer.

**Acknowledgements**

This experiment was carried out under Program Number NP0702-RIBF20 at the RIBF operated by RIKEN Nishina Center, RIKEN and CNS, University of Tokyo. The authors would like to thank the RIBF accelerator crew.  The authors SM, JN were supported by the U.S. Department of Energy, Office of Nuclear Physics, under Contract No. DE-AC02-06CH11357. The authors AN, OT, DB, BS were supported by the National Science Foundation under Grant No. PHY-0606007 and by the U.S. Department of Energy, Office of Nuclear Physics, under Grant No. DE-FG02-03ER41265. The author MF was supported by the National Science Foundation under Grants No. PHY-0855013 and PHY-0735989.

Table 1 Summary of the experimental conditions.

| Setting | G1 (for Z~30) | G2 (for Z~40) | G3 (for Z~50) |
|---|---|---|---|
| $B\rho$ [1) | 7.902 Tm | 7.990 Tm | 7.706 Tm |
| Production target | Be 5.1 mm | Be 2.9 mm | Pb 0.95mm +Al 0.3 mm |
| Degrader at F1 | Al 1.29 mm | Al 2.18 mm | Al 2.56 mm |
| Degrader at F5 | None | None | Al 1.8 mm |
| F1 slit | ±64.2 mm | ±64.2 mm | ±64.2 mm |
| F2 slit | ±13.5 mm | ±15.5 mm | ±15 mm |
| Isotope tuned [2) | $^{79}$Ni | $^{116}$Mo | $^{140}$Sb |
| Average intensity [3) | 0.20 pnA | 0.25 pnA | 0.22 pnA |
| Total dose | $1.17 \times 10^{14}$ particles | $2.35 \times 10^{14}$ particles | $1.34 \times 10^{14}$ particles |
| Average live time | 87% | 94% | 85% |
| Irradiation Time | 25.5 h | 42.5 h | 27.0 h |

1) The values from the magnetic fields of the first dipole magnet.
2) The $B\rho$ setting after F1 is tuned for the isotopes listed.
3) 1 pnA (particle nA) = $6.24 \times 10^{9}$ particles/s.

Table 2 List of the new isotopes identified in the present work.

| Isotope | Setting | Counts | p-value [%] | Cross section [pb] | Isotope | Setting | Counts | p-value [%] | Cross section [pb] | Isotope | Setting | Counts | p-value [%] | Cross section [pb] |
|---|---|---|---|---|---|---|---|---|---|---|---|---|---|---|
| $^{71}$Mn | G1 | 3 | —† | 4 | $^{103}$Rb | G1 | 16 | <0.001 | — | $^{123}$Ru | G2 | 3 | <0.001 | 2 |
| $^{73}$Fe | G1 | 4 | —† | 6 | | G2 | 83 | <0.001 | 110 | $^{124}$Ru | G2 | 1 | <0.001 | 0.6 |
| $^{74}$Fe | G1 | 1 | —† | 1 | $^{106}$Sr | G2 | 22 | <0.001 | 15 | $^{123}$Rh | G2 | 920 | <0.001 | 1470 |
| $^{76}$Co | G1 | 5 | —† | 8 | $^{107}$Sr | G2 | 2 | 0.48 | 1 | | G3 | 11 | <0.001 | — |
| $^{79}$Ni | G1 | 3 | —† | 5 | $^{108}$Y | G1 | 10 | 0.032 | — | $^{124}$Rh | G2 | 94 | <0.001 | 110 |
| $^{81}$Cu | G1 | 36 | <0.001 | 70 | | G2 | 122 | <0.001 | 97 | $^{125}$Rh | G2 | 13 | <0.001 | 11 |
| $^{82}$Cu | G1 | 2 | —† | 3 | $^{109}$Y | G2 | 6 | <0.001 | 4 | $^{126}$Rh | G2 | 1 | 0.46 | 0.7 |
| $^{84}$Zn | G1 | 22 | <0.001 | 43 | $^{111}$Zr | G2 | 26 | <0.001 | 20 | $^{127}$Pd | G2 | 70 | <0.001 | 80 |
| $^{85}$Zn | G1 | 1 | —† | 2 | $^{112}$Zr | G2 | 1 | 0.55 | 7 | | G3 | 1 | 0.53 | — |
| $^{87}$Ga | G1 | 10 | <0.001 | 19 | $^{114}$Nb | G2 | 15 | <0.001 | 11 | $^{128}$Pd | G2 | 13 | <0.001 | 12 |
| $^{90}$Ge | G1 | 3 | 0.0043 | 6 | $^{115}$Nb | G2 | 4 | <0.001 | 3 | $^{133}$Cd | G2 | 11 | <0.001 | 26 |
| $^{95}$Se | G1 | 9 | <0.001 | 20 | $^{115}$Mo | G2 | 933 | <0.001 | 1150 | | G3 | 2 | 0.31 | 40 |
| | G2 | 6 | <0.001 | — | $^{116}$Mo | G2 | 78 | <0.001 | 72 | $^{138}$Sn | G3 | 23 | <0.001 | 600 |
| $^{98}$Br | G1 | 6 | 0.094 | 10 | $^{117}$Mo | G2 | 6 | <0.001 | 4 | $^{140}$Sb | G3 | 124 | <0.001 | 4300 |
| | G2 | 5 | <0.001 | — | $^{119}$Tc | G2 | 27 | <0.001 | 24 | $^{143}$Te | G3 | 8 | <0.001 | 300 |
| $^{101}$Kr | G1 | 5 | <0.001 | 10 | $^{120}$Tc | G2 | 3 | <0.001 | 2 | $^{145}$I | G3 | 57 | <0.001 | 3100 |
| | G2 | 4 | <0.001 | — | $^{121}$Ru | G2 | 143 | <0.001 | 170 | $^{148}$Xe | G3 | 1 | 0.46 | 70 |
| | | | | | $^{122}$Ru | G2 | 15 | <0.001 | 13 | $^{152}$Ba | G3 | 17 | <0.001 | — |

† The p-value is not given, because no events were observed for neighboring Hydrogen-like peaks and the misidentification is not possible. The p-value gives the probability of misidentification of a new isotope (see text).

The significant figures of cross sections are based on statistical errors. We estimate that the cross sections have systematic errors of ~50%, ~40%, and ~30% for the G1, G2, and G3 settings, respectively (see text). Note that the cross sections are not shown for isotopes whose mean position at F2 is located outside the slit opening (see text).

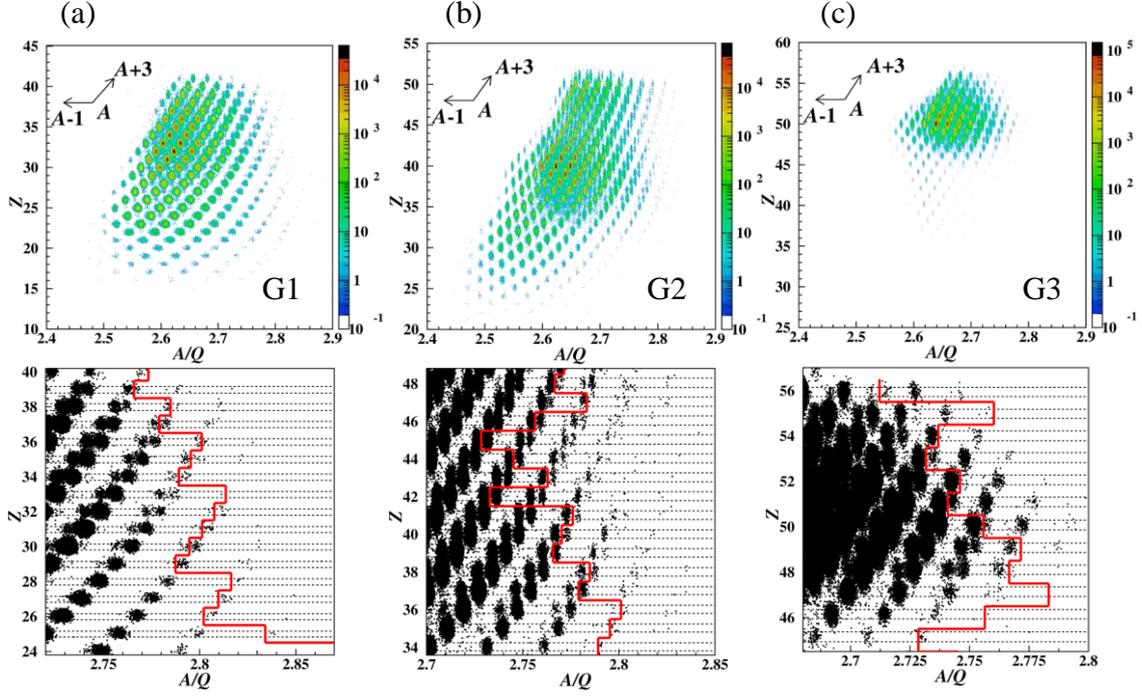

Fig. 1  $Z$ versus $A/Q$ plots for the fission fragments produced in the $^{238}$U+Be reaction (a and b) and the $^{238}$U+Pb reaction (c) at 345 MeV/nucleon. (a) Data obtained with the G1, (b) G2, and (c) G3 settings. The arrows in the upper panels indicate that the isotopes located on the upper right hand and on the left hand correspond to those with mass numbers $A+3$ and $A-1$, respectively. The lower panels show the PID plot enlarged around the regions of new isotopes, where the red lines indicate the known frontier and the dotted horizontal lines show the $Z$ gates.

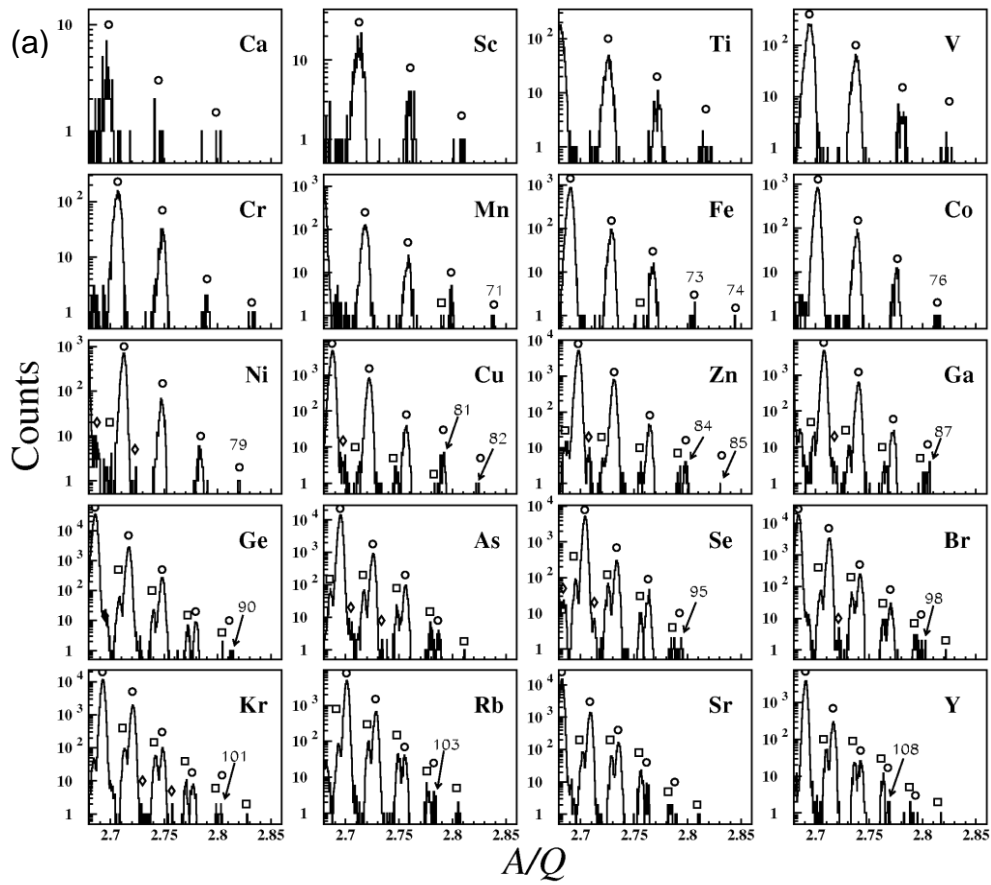
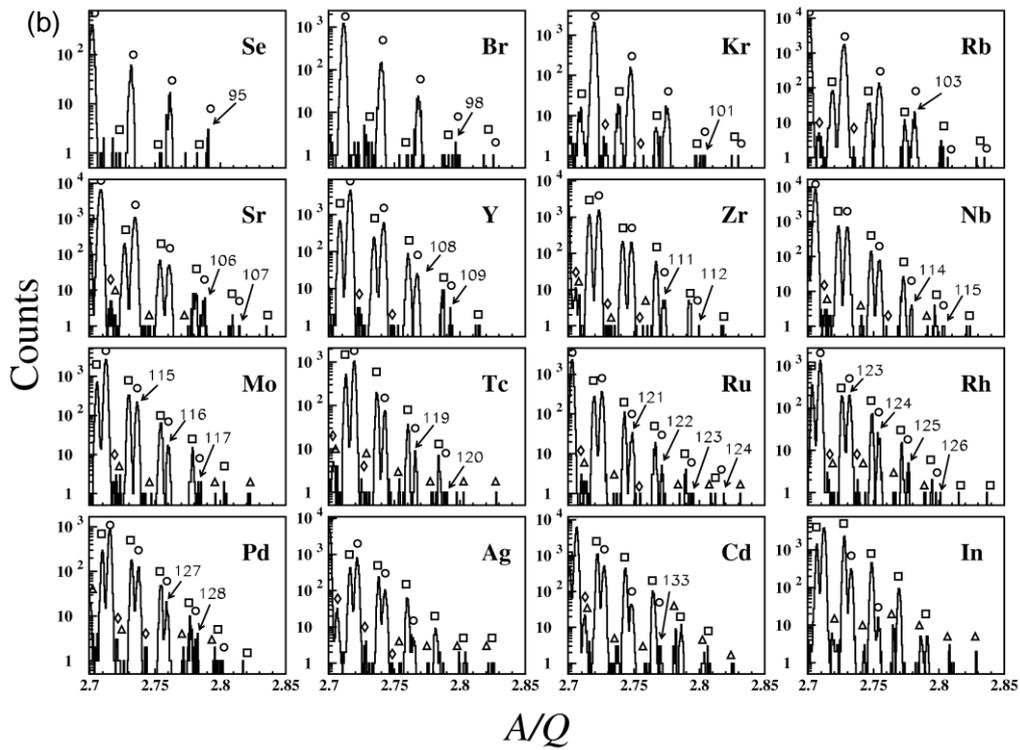

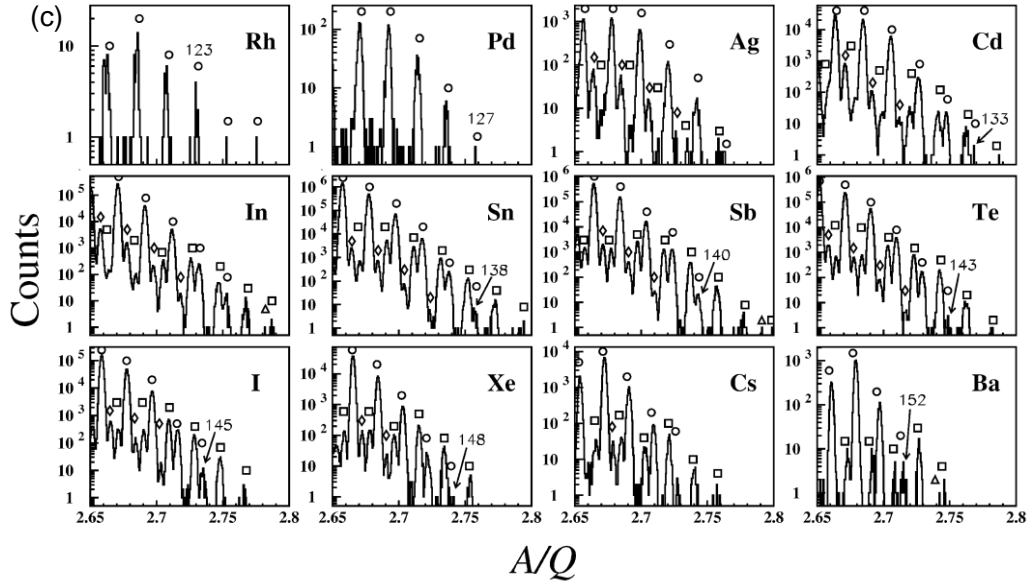

Fig. 2 (a) Shown are *A/Q* spectra of the Ca to Y isotopes (*Z*=20-39) obtained with the G1 setting. (b) Those of the Se to In isotopes (*Z*=34-49) obtained with the G2 setting. (c) Those of the Rh to Ba isotopes (*Z*=45-56) obtained with the G3 setting. The peaks labeled with their mass number correspond to the new isotopes identified in the preset work, while other peaks are labeled by the charge states based on their A/Q values. The charge states *Q=Z*, *Z*-1, and *Z*-2 are indicated by circles, squares, and triangles, respectively. The contamination peaks originated from fully-stripped ions with the neighboring atomic number *Z*+1 are labeled by diamonds.

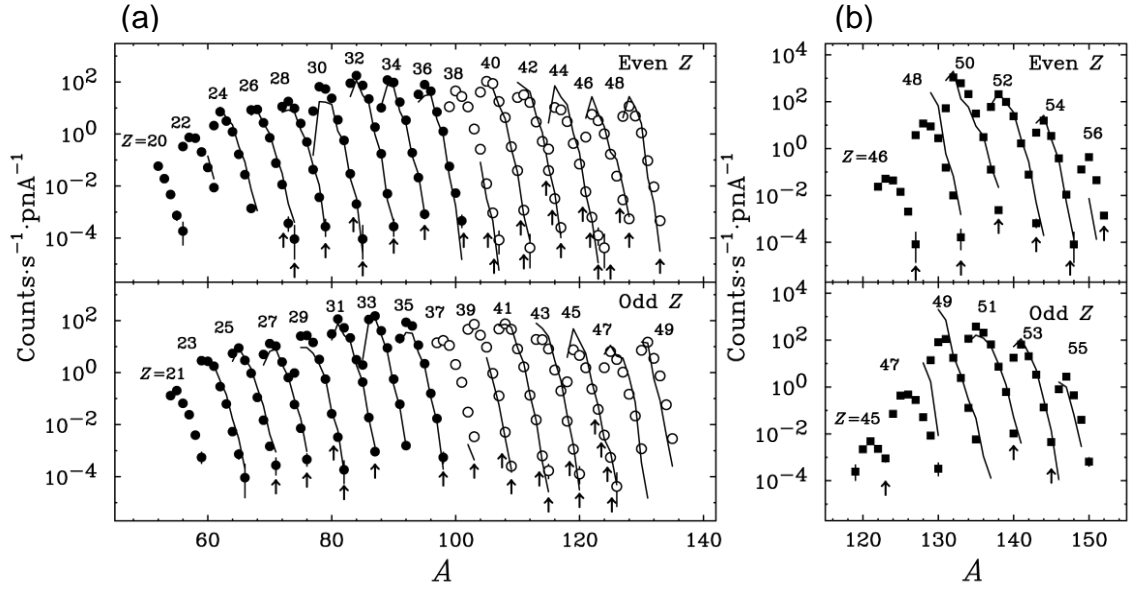

Fig. 3 Measured production rates shown along with the predictions from the LISE++ simulation (solid line) described in the text. (a) The data obtained with the G1 (closed circle) and G2 (open circles) settings, and (b) with the G3 setting. Note that the predictions are not shown for isotopes whose mean position at F2 is located outside the slit opening. The new isotopes are labeled by arrows.

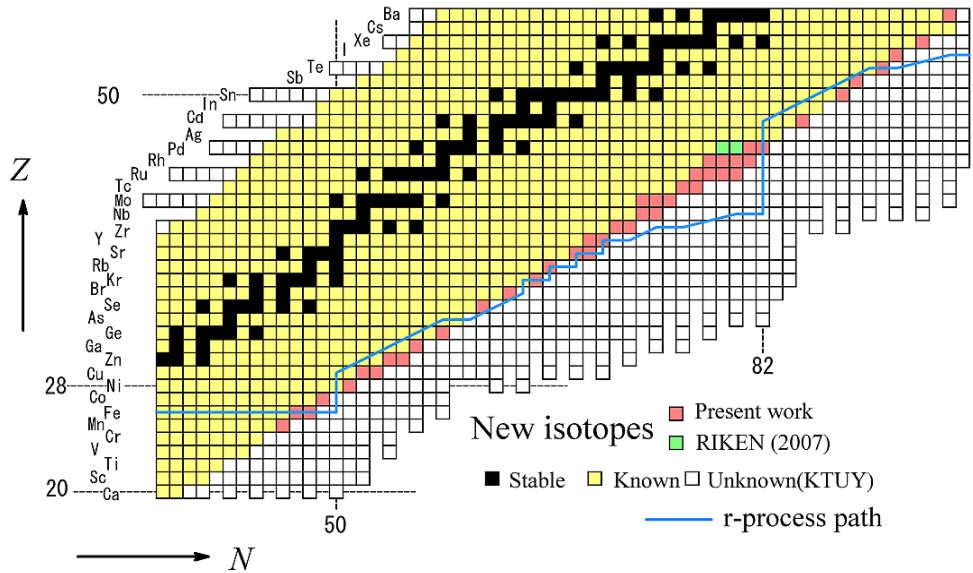

Fig. 4 Nuclear chart based on the KTUY mass model. The new isotopes observed in the present work are shown in red along with an r-process path that is calculated based on the same mass model. The yellow and green squares are previously identified isotopes.